The Evolution of Sociality and the Polyvagal Theory

J. Sean Doody[1,4], Gordon Burghardt[2,3], and Vladimir Dinets[2]

[1]Department of Integrative Biology, University of South Florida – St. Petersburg Campus, 140 7th
Ave. South, St. Petersburg, Florida 33701, USA

[2] Department of Psychology, University of Tennessee, 1404 Circle Drive, Knoxville, TN 37996

[3]Department of Ecology and Evolutionary Biology, University of Tennessee, Knoxville, TN 37996

[4]Corresponding author

Abstract

The polyvagal theory (PT), offered by Porges (2021), proposes that the autonomic nervous system (ANS) was repurposed in mammals, via a "second vagal nerve", to suppress defensive strategies and support the expression of sociality. Three critical assumptions of this theory are that (1) the transition of the ANS was associated with the evolution of 'social' mammals from 'asocial' reptiles; (2) the transition enabled mammals, unlike their reptilian ancestors, to derive a biological benefit from social interactions; and (3) the transition forces a less parsimonious explanation (convergence) for the evolution of social behavior in birds and mammals, since birds evolved from a reptilian lineage. Two recently published reviews, however, provided compelling evidence that the social-asocial dichotomy is overly simplistic, neglects the diversity of vertebrate social systems, impedes our understanding of the evolution of social behavior, and perpetuates the erroneous belief that one group—non-avian reptiles—is incapable of complex social behavior. In the worst case, if PT depends upon a transition from 'asocial reptiles' to 'social mammals,' then the ability of PT to explain the evolution of the mammalian ANS is highly questionable. A great number of social behaviors occur in both reptiles and mammals. In the best case, PT has misused the terms 'social' and 'asocial'. Even here, however, the theory would still need to identify a particular suite of behaviors found in mammals and not reptiles that could be associated with, or explain, the transition of the ANS, and then replace the 'asocial' and 'social' labels with more specific descriptors.

Introduction

The polyvagal theory (PT) posits that there were neurobiological and biobehavioral shifts in the autonomic nervous system (ANS) that occurred during the transition from asocial reptiles to social mammals (Porges, 2021). According to the theory, during this time the ANS was repurposed to suppress defensive strategies to support and express sociality; this 'new ' (our word) mammalian ANS (our words) gave mammals the capacity "…to self-calm, to spontaneous (sic!) socially engage others, and to mitigate threat reactions in ourselves and others through social cues" (Porges, 2021: 1). Anatomically, according to Porges and Dana (2018), a second vagal pathway evolved that had the capacity to down-regulate defense behavior; this second vagal pathway is supposedly found in mammals and not reptiles (Taylor et al., 2022). The adaptive function of this new mammalian ANS was to support homeostatic functions leading to optimized health, growth, and restoration; PT thus, "emphasizes sociality as the core process in mitigating threat reactions and supporting mental and physical health." (Porges, 2021:1).

More specifically, according to Porges (2021), 'the theory focuses on the transition from reptiles to mammals and emphasizes the neural adaptations that enable cues of safety to downregulate states of defense. This transition resulted in the capacity to functionally 'retune' the ANS, thus fostering social engagement behaviors and permitting physiological state co-regulation through social interactions. The theory also claims to explain why mammals have increased metabolic demands compared to non-avian reptiles. The theory emphasizes the need for social interactions in regulating the human ANS and in fostering homeostatic functions.

According to Porges (Porges and Dana, 2018; Porges, 2021), 'the theory further emphasizes that there are unique attributes of the mammalian ANS that differ from reptiles and other earlier vertebrates including the integration of brainstem structures (i.e., ventral vagal complex) to coordinate the regulation of the ventral vagal nucleus (i.e., nucleus ambiguus) with special visceral efferent pathways emerging from cranial nerves V, VII, IX, X, and XI to form an early developmental survival-related suck-swallow-breathe-vocalize circuit. As these pathways mature, they form, as proposed by the (PT), a spontaneous social engagement system that supports homeostasis and co-regulation. Within PT the evolutionary trend has led to a conceptualization of an emergent and uniquely mammalian social engagement system in which a "modified branch of the vagus" is integral. Neuroanatomically, this system is dependent on a brainstem area known as the ventral vagal complex. This area not only regulates the mammalian ventral cardio-inhibitory vagal pathway, but also regulates the special visceral efferent pathways controlling the striated muscles of the face and head.'

Our purpose here is not to critique the neuroanatomical and neurophysiological bases of PT (see Taylor et al., 2022 for such a treatment). Rather, our aim as social behavioral biologists is to address three behavioral assumptions of PT; we find these assumptions controversial at best and scholarly outrageous at worst. Regardless, these assumptions need to be addressed in order for the theory to stand. First, the PT rests the shift in the ANS upon the idea that social mammals evolved from asocial reptiles. Second and relatedly, PT posits that this transition enabled mammals, unlike their reptilian ancestors, to derive a biological benefit from social interactions (Porges 2021). Third, the above claim of an evolutionary transition from asocial reptiles to social mammals forces a less parsimonious explanation (convergence) for the evolution of social behavior in birds and mammals since birds evolved from a reptilian lineage. Given the strength of our critical claims, we expand below our analysis of the legitimacy of these PT assumptions based on the current literature.

Discussion

*Assumption 1. An evolutionary transition from asocial reptiles to social mammals*

Key to PT is that social mammals evolved from asocial reptiles. For example, twice Porges mentions the 'evolutionary transition from asocial reptiles to social mammals,' and then he expounds '…we see shifts in structure and function of the ANS as asocial vertebrates evolved into social mammals with a biological imperative to connect, nurse, cooperate, and trust select others' (Porges, 2021, citing Dobzhansky, 1962).

In reality, mammals did not evolve from reptiles as currently viewed, and, according to current terminology, "reptiles" are no longer a valid phylogenetic term (Prum, 2008). Modern

"reptiles" belong to three lineages, all of which are very distantly related to the ancestors of mammals (Watson, 1957), and two (turtles and particularly crocodilians) are more closely related to birds than to the third lineage, squamates (which includes lizards, snakes, and the Tuatara) (Heilmann, 1927; Iwabe et al., 2004).

Contrary to popular misconception, reptiles can be very social, and although they do not nurse, all of them socially interact, often in complex ways, which are now known to include cases of communal breeding, complex and prolonged parental care, life-long monogamy, extended families with permanent social bonds, complex communication and mating systems, coordinated hunting, coordinated group movements, social learning, social play, social and even self-recognition (c.f., Burghardt et al., 2021), and most other forms of social interactions known for mammals (Fig. 1; see more examples in Doody et al., 2021; Font et al., 2023).

Mammals, on the other hand, are on average much less social than people may think. Mammals most familiar to the general public, such as large primates, dolphins, large ungulates and large canines, are in fact atypical representatives of the class Mammalia. More than 3/4 of all mammalian species are rodents, bats, shrews, small marsupials, small carnivores, and other groups where the overwhelming majority of species are no more social than most reptiles, except for nursing behavior (Nowak and Walker, 1999). The second largest group (after rodents) is bats, which often form large roosting aggregations, but tend to forage alone and have rather simple sociality, with only a few exceptions (Graham, 2001). Even among primates, ungulates and cetaceans, there appear to be taxa that are no more social than a typical viper or crocodile, such as mouse lemurs, duikers, and rorquals (e.g., Brattstrom, 1973). The extent of parental care in mammals is also highly variable: in some species such as the hooded seal it lasts for only a few days (Stewart, 2014) and guinea pigs are so precocial they can eat solid food right after birth, that is, nursing and mother's milk are not even required (Harknes et al., 1995). Of particular importance for the present discussion is the distribution of complex sociality (such as permanent groupings with social hierarchy, communal parental care, coordinated hunting, etc.) among mammalian lineages: it has evolved multiple times in various mammalian orders but does not seem to be the ancestral condition for any one of them (Fig. 1; see also Kutsukake, 2009).

PT is not alone in creating this incorrect dichotomy. Indeed, because social behavior in vertebrates spans a continuum from solitary to highly social, taxa are sometimes dichotomized as either 'social' or 'non-social' (reviewed in Doody et al, 2013; 2021). Two recently published reviews provided compelling evidence that this dichotomy is overly simplistic, neglects the diversity of vertebrate social systems, impedes our understanding of the evolution of social behavior, and perpetuates the erroneous belief that one group—the reptiles—is primarily asocial (Doody et al., 2013; 2021). Regarding reptiles as asocial ignores their diverse and often complex social behaviors, resulting in reptiles being overlooked by researchers examining vertebrate social behavior, limits the scope of the reptile social behavior studies, and creates the impression that reptile species provide little opportunity for studying the mechanisms underlying the evolution of complex social behavior (Doody et al., 2013; 2021). In reality, the types and extent of sociality differ within all of the classes of vertebrates, including reptiles (Doody et al., 2013). Moreover, the effects of this dismissal of reptiles as having complex social behavior, emotions, and even play have been ensconced in the literature by leading neuroscientists based on brain structures (e.g., MacLean, 1985; 1990) and has pernicious effects on reptilian behavioral research (Burghardt, 2020). However, recent comparative neuroscience is documenting the ancient evolutionary origins of neural structures underlying complex social behavior (e. g. Bass and Chagnaud, 2012; Font et al., 2023). Finally, the social repertoire of the often more-secretive reptiles has been little-studied, as compared to mammals and birds (Doody et al., 2021), leading to more uncertainty as to where reptiles stand with other vertebrates and animals within a comparative social context.

A better approach would recognize a continuum along which each species sits that ranges from weakly to strongly social (Doody et al., 2021). At the strongly social end of the continuum sits obligatory eusociality: caste systems involving divisions of labor in reproduction, defense, and foraging, overlapping generations maintaining prolonged contact with each other, and cooperative care of young. At the weakly social end sit animals that rarely interact with conspecifics, other than during brief, simple mating encounters, and of course solitary asexual animals (Doody et al., 2021), represented among reptiles by only a handful of species (Dubach et al., 1997). Deciding where a species sits on this continuum is a difficult task; this continuum is not to be confused with the solitary to group-living continuum, in spite of frequent overlap; levels and types of sociality are more like a tree than a linear system (Doody et al., 2021).

A consensus framework that adequately classifies the magnitude of social behavior among all animal taxa has not been developed (but see Kappeler, 2019). Michener (1969) developed a classification for social insects that included 'solitary' (no parental care), 'subsocial' (adults care for their own young), 'communal' (adults use the same nest but do not care for the brood cooperatively), 'quasisocial' (cooperative brood care in the same nest), semisocial (quasisocial plus a worker caste), and eusocial (semisocial plus an overlap in adult generations). While this framework distinguishes between processes that allows evolutionary insights, it does not address variation among animals with different social repertoires. For example, even among species without parental care after birth or hatching, there is considerable variation in the presence of other social behaviors such as dominance hierarchies, territoriality, male-to-male combat, complex courtship, group vigilance, signaling, posturing, eavesdropping, communal nesting, cooperative hunting, pair bonding, sexual selection, and social monogamy (all of which occur in reptiles; Doody et al., 2021). Similarly, there have been attempts (Chance and Jolly, 1970; Crook, 1970) to classify non-human primate societies based on mating systems that include a mother an offspring (e.g., orangutans), polygyny harems (e.g., some baboons), monogamy (e.g., gibbons), polygamy (e.g., many macaque monkeys), polyandry (e.g., marmosets, tamarins), and a fission-fusion society (e.g., chimpanzees). These are further embedded in hierarchical or territorial systems that may be seasonal or permanent. Several of these systems are found in reptiles (Doody et al., 2021).

Replacing a one-dimensional scale with a multidimensional one would be an improvement (Kappeler, 2019) but problems remain. A framework that adequately facilitates the classification of all animals would require a broad perspective and intensive review, and would be subject to much debate and disagreement as to which animals fit where on a continuum. Are polygamous species more social than monogamous ones? Are those with dominance hierarchies more social than territorial species? Are large herding species more social than their group-hunting predators? Simple metrics are hard to apply across disparate taxa. Kappeler (2019) offered an encouraging framework with four core components of a social system: social organization, social structure, mating system, and care system. This paper will likely stir a debate leading to improved frameworks, concepts and definitions that will facilitate further study of the evolution of social behavior among vertebrates and other animals (see also Rubenstein and Abbot, 2017).

Returning to PT, Porges' dichotomy is incorrect. While many mammals (particularly humans) may show more complex social behavior than reptiles, there is considerable overlap in social tendencies between the two groups. The labels 'social' and 'asocial' are too crude to have utility in a comparative framework of social behavior and should not be used to describe taxa, especially within the context of explaining a key innovation. Moreover, reptiles are not asocial, although even reptile researchers who should know better continue to make such claims that are then repeated in the literature. They span a wide range of social tendencies, and some of them have been highly social even hundreds of millions of years ago; for example, dinosaurs had complex parental care, communal nesting, and gregarious foraging (see, for example, Horner, 2000; Varicchio et al., 2008; Padian et al., 2011; Fiorillo et al., 2014; Pol et al., 2021). Sociality might even be the ancestral condition of amniotes, since their extant sister group, the modern amphibians, is largely composed of highly social species. Although there is remarkably diverse paleontological evidence of social behavior in extinct reptiles (reviewed in Doody et al., 2021), there is only one case of such evidence for the amniote lineage leading to mammals (Botha-Brink and Modesto, 2007) and none for early mammals. We do not yet understand how social behavior evolved from early amniotes to reptiles, mammals and birds. We do know that it is extremely unlikely that there was a neurological switch within a stem mammal (via PT) that became a key innovation for mammals at the exclusion of other vertebrates. If there was such a switch or repurposing of the ANS, it cannot be associated with a dichotomous branching of ancient 'asocial' reptiles and 'social' mammals. We know this because the distribution of social behaviors in reptiles and mammals does not support the social-asocial dichotomy (Doody et al., 2013, 2021). Moreover, the genes underpinning social behaviors would likely be present in early amniotes. PT thus, appears to reflect a naïve view of the neuroanatomical basis of social behavior.

*Assumption 2. Reptiles do not derive a biological benefit from social interactions.*

A related claim by Porges (2021) is that the 'transition (of the ANS) enabled mammals, unlike their reptilian ancestors, to derive a biological benefit from social interactions." Above we outlined the many social behaviors exhibited by reptiles. It would be absurd to consider that these social behaviors evolved without providing a biological benefit. Social behaviors exhibited by reptiles surely provide benefits to individuals of those species, including group living (e.g., some skinks), monogamy (e.g., sleepy lizards), nest guarding (e.g., many lizards, virtually all crocodilians, some snakes, a few turtles, the tuatara), crèche behavior (e.g., gharials), communal nesting (e.g., many lizards, turtles, some snakes and crocodilians, tuatara), territoriality (e.g., most lizards) and social learning (e.g., turtles, lizards) (see Doody et al., 2021). Indeed, these behaviors occur in other vertebrates including mammals and all would have at least provided benefits when they evolved and likely provide them today. Perhaps what Porges meant was that the 'self-calming homeostasis' that he claims is restricted to mammals via PT is a biological benefit that reptiles did not receive, in which case we would refer to assumption 1. Or, perhaps, he was referring to reduced stress levels in group-forming animals, but this is contradicted by some studies (for example, Rodenburg and Koene, 2003; see also an overview in Blanchard et al., 2001), and even if such an effect does exist, there is no reason to think that it exists in mammals but not in group-forming reptiles such as many crocodilians, lizards and turtles. In any case, it is difficult to imagine that an obsessively vigilant and nervous mammal such as a chevrotain (mouse deer) has better "self-calming abilities" than, for example, a dwarf chameleon that exhibits no visible behavioral reaction to being caught and handled (VD, pers. obs.).

It is also possible that Porges meant to say was that mammals have somehow reached some threshold of complex social behavior that is currently out of reach for reptiles, although he did not articulate this. In the last section we highlighted the lack of a consensus for assigning a social score to a species and thus a group, so even if the theory was referring to some threshold, this has not been established. Moreover, there are clearly continuums of social traits within both mammals and reptiles along which arbitrary lines cannot be drawn for understanding PT and the evolution of social behavior. Another possibility is that Porges was referring only to parental care - less than 3% of reptiles engage in post-ovipositional parental care (Doody et al., 2013). Porges (2021) claims that PT does not focus on the obvious similarities with more ancient vertebrates, but rather, it focuses on the 'unique modifications that enabled mammals to optimize their survival.' However, his list included 'connecting, cooperating and trusting select others' (Porges, 2021), all of which are social behaviors that are not restricted to parental care, or to mammals.

*Assumption 3. Social behavior in birds creates a problem with the above assumptions*

According to Porges (2021), since PT is focused on the phylogenetic transition from reptiles to mammals, it does not focus on the sociality of other vertebrate species that evolved from reptiles (e.g., birds) following the divergence of mammals. However, PT theory, with mammals as distinctly social, forces a less parsimonious explanation (more convergence) for the evolution of social behavior in birds and mammals, since birds evolved from a reptilian lineage. For example, by mapping the known reproductive behaviors of non-avian dinosaurs on a cladogram, including the social behaviors of nest attendance, parental protection, brooding, and communal nesting, Horner (2000) revealed that many of the reproductive traits of birds may have been derived from both their archosaurian and dinosaurian ancestors (see also Varricchio et al. 2008, arguing that parental care in birds has its origin in dinosaurs). Similarly, mapping the earliest evidence of group courtship, communal breeding and social hunting/foraging on a cladogram of amniotic vertebrates reveals complexity that cannot be reconciled with PT (Fig. 1). Although mammals evolved nursing and lactation, a great number of other social behaviors have occurred in the ancestors of reptiles (and thus, birds) – for example, group living, communal nesting/birthing, courtship and mating, nest guarding, territoriality, crèche behavior and social learning, with the possibility of other behaviors such as monogamy, group defense, and cooperative breeding in dinosaurs and other extinct reptilian lineages. Many birds feed offspring half-digested food, and crop 'milk' in pigeons, flamingoes and some penguins is a well-known phenomenon (Eraud et al., 2008), and it has been proposed that only something similar can explain extremely high growth rates in sauropod dinosaurs (Erickson et al., 2001). Maternal feeding of hatchling caecilians from skin glands has also been reported (Wilkinson, et al. 2008), so lactation-like behavior is not a mammalian invention.

*Summary*

In the worst case, if PT depends upon a postulated transition from 'asocial reptiles' to 'social mammals' then it is demonstrably false. These labels do not just involve semantics. Sociality is as diverse, and likely more ancient, in reptiles compared to mammals. Although there is a small number of social behaviors that occur in non-human mammals and not in extant reptiles (e.g., nursing, cooperative breeding, group warfare), Porges (Porges and Dana, 2018; Porges, 2021) did not specifically identify these as critical to the proposed transition. In the best case, PT has misused the terms 'social' and 'asocial'. In this case, however, PT would still need to identify a particular behavior or suite of behaviors found in mammals and not reptiles that could be associated with, or explain, the transition of the ANS, and then replace the 'asocial' and 'social' labels with more specific descriptors. As currently presented, PT appears to rest upon 20th century folk interpretation of vertebrate evolutionary biology rather than on current scientific understanding of it.

## Acknowledgments

We thank P. Grossman for inviting us to share our perspective.

Figure caption:

Fig. 1. Cladogram of amniotic vertebrates (with some extinct lineages omitted). Arrows show the earliest evidence of social behavior: empty arrows - parental care; filled arrows - group courtship, communal breeding, social hunting/foraging). Note that in many clades social behavior is clearly much older than suggested by fossil evidence or lack thereof. Mln=Millions of years ago. Modified from Doody et al. (2021).

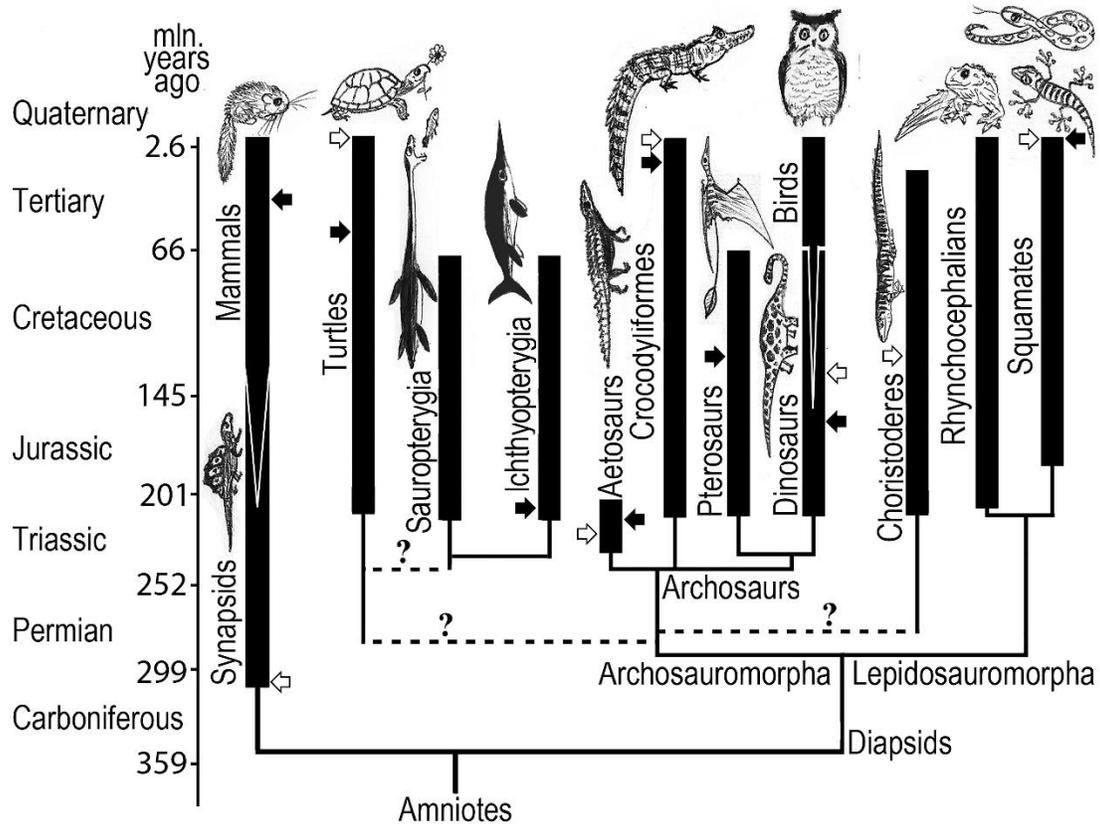